\pdfoutput=1
\documentclass[12pt,a4paper]{article}
\usepackage{ifthen} % for conditional statements
\newboolean{pdflatex}
\setboolean{pdflatex}{true} % False for eps figures 

\newboolean{articletitles}
\setboolean{articletitles}{true} % False removes titles in references

\newboolean{uprightparticles}
\setboolean{uprightparticles}{false} %True for upright particle symbols

\usepackage{microtype}
\usepackage{lineno}  % for line numbering during review
\usepackage{xspace} % To avoid problems with missing or double spaces after
\usepackage{graphicx}  % to include figures (can also use other packages)
\usepackage{color}
\usepackage{colortbl}
\graphicspath{{./figs/}} % Make Latex search fig subdir for figures

\usepackage{amsmath} % Adds a large collection of math symbols
\usepackage{amssymb}
\usepackage{amsfonts}
\usepackage{upgreek} % Adds in support for greek letters in roman typeset

\newcommand*\patchAmsMathEnvironmentForLineno[1]{%
\expandafter\let\csname old#1\expandafter\endcsname\csname #1\endcsname
\expandafter\let\csname oldend#1\expandafter\endcsname\csname
end#1\endcsname
 \renewenvironment{#1}%
   {\linenomath\csname old#1\endcsname}%
   {\csname oldend#1\endcsname\endlinenomath}%
}
\newcommand*\patchBothAmsMathEnvironmentsForLineno[1]{%
  \patchAmsMathEnvironmentForLineno{#1}%
  \patchAmsMathEnvironmentForLineno{#1*}%
}
\AtBeginDocument{%
\patchBothAmsMathEnvironmentsForLineno{equation}%
\patchBothAmsMathEnvironmentsForLineno{align}%
\patchBothAmsMathEnvironmentsForLineno{flalign}%
\patchBothAmsMathEnvironmentsForLineno{alignat}%
\patchBothAmsMathEnvironmentsForLineno{gather}%
\patchBothAmsMathEnvironmentsForLineno{multline}%
}

\usepackage{hyperref}    % Hyperlinks in references
\usepackage[all]{hypcap} % Internal hyperlinks to floats.

\def\lhcb {\mbox{LHCb}\xspace}
\def\ux85 {\mbox{UX85}\xspace}

\def\lhc    {\mbox{LHC}\xspace}

\ifthenelse{\boolean{uprightparticles}}%
{

 \def\Pmu         {\ensuremath{\upmu}\xspace}

 \def\Ppi         {\ensuremath{\uppi}\xspace}

 \def\PDelta      {\ensuremath{\Delta}\xspace}                 
 \def\PXi      {\ensuremath{\Xi}\xspace}                 
 \def\PLambda      {\ensuremath{\Lambda}\xspace}                 
 \def\PSigma      {\ensuremath{\Sigma}\xspace}                 
 \def\POmega      {\ensuremath{\Omega}\xspace}                 
 \def\PUpsilon      {\ensuremath{\Upsilon}\xspace}                 
 
 \def\PB      {\ensuremath{\mathrm{B}}\xspace}                 
                  
 \def\PD      {\ensuremath{\mathrm{D}}\xspace}

 \def\PK      {\ensuremath{\mathrm{K}}\xspace}

 \def\Pb      {\ensuremath{\mathrm{b}}\xspace}                 
 \def\Pc      {\ensuremath{\mathrm{c}}\xspace}

 \def\Pi      {\ensuremath{\mathrm{i}}\xspace}

 \def\Ps      {\ensuremath{\mathrm{s}}\xspace}

}
{

 \def\Pmu         {\ensuremath{\mu}\xspace}

 \def\Ppi         {\ensuremath{\pi}\xspace}

 \mathchardef\PDelta="7101
 \mathchardef\PXi="7104
 \mathchardef\PLambda="7103
 \mathchardef\PSigma="7106
 \mathchardef\POmega="710A
 \mathchardef\PUpsilon="7107
                  
 \def\PB      {\ensuremath{B}\xspace}                 
                  
 \def\PD      {\ensuremath{D}\xspace}

 \def\PK      {\ensuremath{K}\xspace}

 \def\Pb      {\ensuremath{b}\xspace}                 
 \def\Pc      {\ensuremath{c}\xspace}

 \def\Pi      {\ensuremath{i}\xspace}

 \def\Ps      {\ensuremath{s}\xspace}

}

\def\mup        {\ensuremath{\Pmu^+}\xspace}
\def\mun        {\ensuremath{\Pmu^-}\xspace} % muon negative (\mum is taken)

\def\squark    {\ensuremath{\Ps}\xspace}

\def\cquark    {\ensuremath{\Pc}\xspace}

\def\bquark    {\ensuremath{\Pb}\xspace}

\def\pion  {\ensuremath{\Ppi}\xspace}

\def\pip   {\ensuremath{\pion^+}\xspace}

\def\kaon  {\ensuremath{\PK}\xspace}
  \def\Kbar  {\kern 0.2em\overline{\kern -0.2em \PK}{}\xspace}

\def\Kz    {\ensuremath{\kaon^0}\xspace}
\def\Kzb   {\ensuremath{\Kbar^0}\xspace}
\def\KzKzb {\ensuremath{\Kz \kern -0.16em \Kzb}\xspace}
\def\Kp    {\ensuremath{\kaon^+}\xspace}
\def\Km    {\ensuremath{\kaon^-}\xspace}

\def\KpKm  {\ensuremath{\Kp \kern -0.16em \Km}\xspace}

  \def\Dbar    {\kern 0.2em\overline{\kern -0.2em \PD}{}\xspace}
\def\D       {\ensuremath{\PD}\xspace}

\def\Dz      {\ensuremath{\D^0}\xspace}
\def\Dzb     {\ensuremath{\Dbar^0}\xspace}
\def\DzDzb   {\ensuremath{\Dz {\kern -0.16em \Dzb}}\xspace}
\def\Dp      {\ensuremath{\D^+}\xspace}
\def\Dm      {\ensuremath{\D^-}\xspace}

\def\DpDm    {\ensuremath{\Dp {\kern -0.16em \Dm}}\xspace}

\def\Dstarm  {\ensuremath{\D^{*-}}\xspace}

\def\Dsm     {\ensuremath{\D^-_\squark}\xspace}

\def\Dssm    {\ensuremath{\D^{*-}_\squark}\xspace}

\def\B       {\ensuremath{\PB}\xspace}
  \def\Bbar    {\kern 0.18em\overline{\kern -0.18em \PB}{}\xspace}

\def\Bz      {\ensuremath{\B^0}\xspace}

\def\Bd      {\ensuremath{\B^0}\xspace}
\def\Bs      {\ensuremath{\B^0_\squark}\xspace}

  \def\Y#1S{\ensuremath{\PUpsilon{(#1S)}}\xspace}% no space before {...}!

\def\L {\ensuremath{\PLambda}\xspace}
\def\Lbar {\ensuremath{\kern 0.1em\overline{\kern -0.1em\PLambda}}\xspace}

\def\Lb      {\ensuremath{\L^0_\bquark}\xspace}
\def\Lbbar   {\ensuremath{\Lbar^0_\bquark}\xspace}
\def\Lc      {\ensuremath{\L^+_\cquark}\xspace}

\def\BF         {{\ensuremath{\cal B}\xspace}}

\def\BR         {\BF}
\newcommand{\decay}[2]{\ensuremath{#1\!\to #2}\xspace}         % {\Pa}{\Pb \Pc}

\def\to                 {\ensuremath{\rightarrow}\xspace}

\def\AT#1     {\ensuremath{A_{\mathrm{T}}^{#1}}\xspace}           % 2

\def\Bsmm     {\decay{\Bs}{\mup\mun}}

\def\C#1      {\ensuremath{\mathcal{C}_{#1}}\xspace}                       % 9
\def\Cp#1     {\ensuremath{\mathcal{C}_{#1}^{'}}\xspace}                    % 7
\def\Ceff#1   {\ensuremath{\mathcal{C}_{#1}^{\mathrm{(eff)}}}\xspace}        % 9  
\def\Cpeff#1  {\ensuremath{\mathcal{C}_{#1}^{'\mathrm{(eff)}}}\xspace}       % 7
\def\Ope#1    {\ensuremath{\mathcal{O}_{#1}}\xspace}                       % 2
\def\Opep#1   {\ensuremath{\mathcal{O}_{#1}^{'}}\xspace}                    % 7

\newcommand{\tev}{\ensuremath{\mathrm{\,Te\kern -0.1em V}}\xspace}
\newcommand{\gev}{\ensuremath{\mathrm{\,Ge\kern -0.1em V}}\xspace}
\newcommand{\mev}{\ensuremath{\mathrm{\,Me\kern -0.1em V}}\xspace}
\newcommand{\kev}{\ensuremath{\mathrm{\,ke\kern -0.1em V}}\xspace}
\newcommand{\ev}{\ensuremath{\mathrm{\,e\kern -0.1em V}}\xspace}
\newcommand{\gevc}{\ensuremath{{\mathrm{\,Ge\kern -0.1em V\!/}c}}\xspace}
\newcommand{\mevc}{\ensuremath{{\mathrm{\,Me\kern -0.1em V\!/}c}}\xspace}
\newcommand{\gevcc}{\ensuremath{{\mathrm{\,Ge\kern -0.1em V\!/}c^2}}\xspace}
\newcommand{\gevgevcccc}{\ensuremath{{\mathrm{\,Ge\kern -0.1em V^2\!/}c^4}}\xspace}
\newcommand{\mevcc}{\ensuremath{{\mathrm{\,Me\kern -0.1em V\!/}c^2}}\xspace}

\def\mum  {\ensuremath{\,\upmu\rm m}\xspace}

\def\fb   {\ensuremath{\mbox{\,fb}}\xspace}

\newcommand{\chisq}{\ensuremath{\chi^2}\xspace}

\def\gsim{{~\raise.15em\hbox{$>$}\kern-.85em
          \lower.35em\hbox{$\sim$}~}\xspace}
\def\lsim{{~\raise.15em\hbox{$<$}\kern-.85em
          \lower.35em\hbox{$\sim$}~}\xspace}

\def\pt         {\mbox{$p_{\rm T}$}\xspace}

\def\evtgen     {\mbox{\textsc{EvtGen}}\xspace}
\def\pythia     {\mbox{\textsc{Pythia}}\xspace}

\def\geant      {\mbox{\textsc{Geant4}}\xspace}

\def\photos     {\mbox{\textsc{Photos}}\xspace}

\def\tell1  {TELL1\xspace}
\def\ukl1   {UKL1\xspace}

\usepackage{cite} % Allows for ranges in citations
\usepackage{mciteplus}

\begin{document}

%%%%%%%%%%%%%%%%%%%%%%%%%
%%%%% Title     %%%%%%%%%
%%%%%%%%%%%%%%%%%%%%%%%%%
\newcommand\TVA{\rule{0pt}{3.6ex}}
\newcommand\BVA{\rule[-2.2ex]{0pt}{0pt}}

\newcommand{\DdK}{\ensuremath{\Dm K^+}}
\newcommand{\Ddp}{\ensuremath{\Dm \pi^+}}

\newcommand{\BdDp}{\texorpdfstring{\decay{\Bz}{\Dm \pip}}{}}
\newcommand{\BdDK}{\texorpdfstring{\decay{\Bz}{\Dm \Kp}}{}}
\newcommand{\BsDp}{\texorpdfstring{\decay{\Bs}{\Dsm \pip}}{}}

\newcommand{\BdDsp}{\ensuremath{\Bd      \to \Dsm       \pi^+\,}}
\newcommand{\LbDsp}{\texorpdfstring{\decay{\Lb}{\Dsm p}}{}}
\newcommand{\LbDsstp}{\decay{\Lb}{\Dssm p}}
\newcommand{\BdDstarp}{\ensuremath{\Bd  \to \Dstarm    \pi^+\,}}
\newcommand{\BdDstarK}{\ensuremath{\Bd  \to \Dstarm      K^+\,}}
\newcommand{\BdDKst}{\ensuremath{\Bd \to \Dm K^{*+} \,}}
\newcommand{\BdDrho}{\ensuremath{\Bd  \to \Dm      \rho^+\,}}
\newcommand{\LbLcp}{\ensuremath{\Lb      \to \Lc \pi^-\,}}
\newcommand{\LbbarLcp}{\ensuremath{\Lbbar \to \L^-_\cquark \pi^+}}
\newcommand{\LbLcK}{\ensuremath{\Lb      \to \Lc K^-\,}}
\newcommand{\LbbarLcK}{\ensuremath{\Lbbar      \to \L^-_\cquark K^+\,}}
\newcommand{\BsDstarp}{\ensuremath{\Bs  \to \Dssm    \pi^+\,}}
\newcommand{\BsDstarK}{\ensuremath{\Bs  \to \Dssm      K^+\,}}
\newcommand{\BsDrho}{\ensuremath{\Bs  \to \Dsm      \rho^+\,}}
\newcommand{\BsDppp}{\ensuremath{\Bs  \to D_s^{-} \pi^+ \pi^{+} \pi^{-}}}

\newcommand{\fdfs}{\ensuremath{\frac{f_d}{f_s}}}
\newcommand{\fsfd}{\ensuremath{\frac{f_s}{f_d}}}
\newcommand{\fsfdt}{\texorpdfstring{\ensuremath{f_s/f_d}~}{}}
\newcommand{\fsfdtns}{\texorpdfstring{\ensuremath{f_s/f_d}}{}}

\renewcommand{\thefootnote}{\fnsymbol{footnote}}
\setcounter{footnote}{1}

% $Id: title-LHCb-PAPER.tex 30318 2013-01-22 16:45:35Z bstoraci $
% ===============================================================================
% Purpose: LHCb-PAPER journal paper title page template
% Author: 
% Created on: 2010-09-25
% ===============================================================================

%%%%%%%%%%%%%%%%%%%%%%%%%
%%%%%  TITLE PAGE  %%%%%%
%%%%%%%%%%%%%%%%%%%%%%%%%
\begin{titlepage}
\pagenumbering{roman}

% Header ---------------------------------------------------
\vspace*{-1.5cm}
\centerline{\large EUROPEAN ORGANIZATION FOR NUCLEAR RESEARCH (CERN)}
\vspace*{1.5cm}
\hspace*{-0.5cm}
\begin{tabular*}{\linewidth}{lc@{\extracolsep{\fill}}r}
\ifthenelse{\boolean{pdflatex}}% Logo format choice
{\vspace*{-2.7cm}\mbox{\!\!\!\includegraphics[width=.14\textwidth]{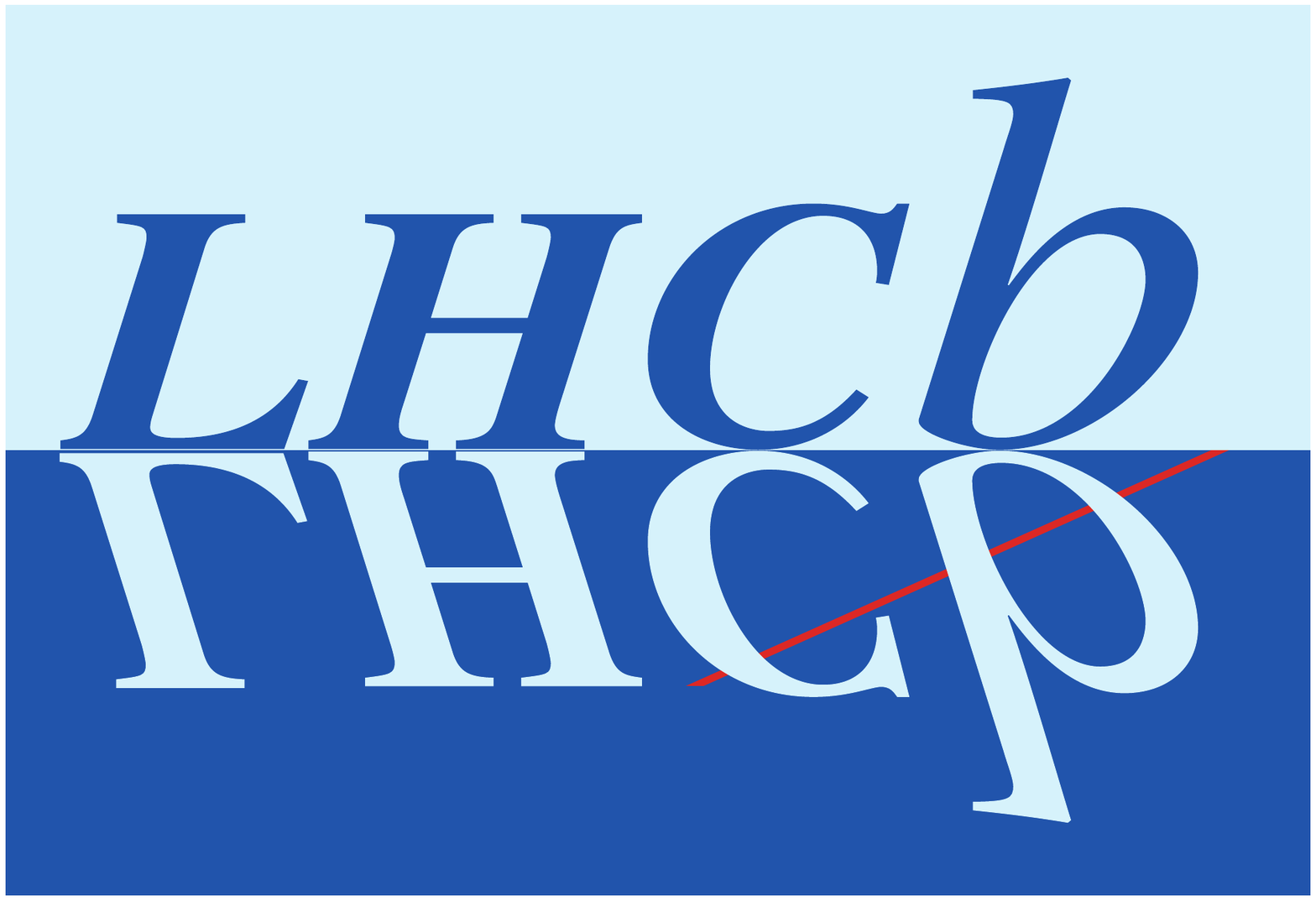}} & &}%
{\vspace*{-1.2cm}\mbox{\!\!\!\includegraphics[width=.12\textwidth]{lhcb-logo.eps}} & &}%
\\
 & & CERN-PH-EP-2013-006 \\  % ID 
 & & LHCb-PAPER-2012-037 \\  % ID 
 & & 22 January 2013 \\ % Date - Can also hardwire e.g.: 23 March 2010
 & & \\
% not in paper \hline
\end{tabular*}

\vspace*{4.0cm}

% Title --------------------------------------------------
{\bf\boldmath\huge
\begin{center}
Measurement of the fragmentation fraction ratio $f_{s}/f_{d}$ and its dependence on $B$ meson kinematics
\end{center}
}

\vspace*{1.0cm}

% Authors -------------------------------------------------
\begin{center}
The LHCb collaboration\footnote{Authors are listed on the following pages.}
\end{center}

\vspace{\fill}

% Abstract -----------------------------------------------
\begin{abstract}
  \noindent
 The relative production rate of $B^{0}_{s}$ and $B^{0}$ mesons is determined
  with the hadronic decays $B^{0}_{s} \rightarrow D^{-}_{s}\pi^{+}$
  and $B^0 \rightarrow D^{-}K^{+}$. The measurement uses data corresponding to
  1.0~fb$^{-1}$ of $pp$ collisions at a centre-of-mass energy of
  $\sqrt{s}=7$~TeV recorded in the forward region with the LHCb
  experiment. The ratio of production rates, $f_{s}/f_{d}$, is measured
  to be $0.238 \pm 0.004 \pm 0.015 \pm 0.021 $, where the first
  uncertainty is statistical, the second systematic, and the third
  theoretical. This is combined with a previous LHCb measurement to
  obtain $\fsfdt = 0.256 \pm 0.020$. The dependence of $f_{s}/f_{d}$
  on the transverse momentum and pseudorapidity of the $B$ meson is
  determined using the decays $B^{0}_{s} \rightarrow D^{-}_{s}\pi^{+}$
  and $B^{0} \rightarrow D^{-}\pi^{+}$. There is evidence for a
  decrease with increasing transverse momentum, whereas the ratio
  remains constant as a function of pseudorapidity. In addition, the
  ratio of branching fractions of the decays \mbox{$B^{0} \rightarrow
  D^{-}K^{+}$} and $B^{0} \rightarrow D^{-}\pi^{+}$ is measured to be
  \mbox{$0.0822 \pm 0.0011\,(\textrm{stat}) \pm 0.0025\,(\textrm{syst})$}.
\end{abstract}

\vspace*{0.0cm}

\begin{center}
  Submitted to Journal of High Energy Physics
\end{center}

\vspace{\fill}

{\footnotesize 
\centerline{\copyright~CERN on behalf of the \lhcb collaboration, license \href{http://creativecommons.org/licenses/by/3.0/}{CC-BY-3.0}.}}

\end{titlepage}

%%%%%%%%%%%%%%%%%%%%%%%%%%%%%%%%
%%%%%  EOD OF TITLE PAGE  %%%%%%
%%%%%%%%%%%%%%%%%%%%%%%%%%%%%%%%

%  empty page follows the title page ----
\newpage
\setcounter{page}{2}
\mbox{~}
\newpage

% Author List ----------------------------
%  You need to get a new author list!
%%%%%%%%%%%%%%%%%%%%%%%%%%%%%%%%%%%%%%%%%%
\centerline{\large\bf LHCb collaboration}
\begin{flushleft}
\small
R.~Aaij$^{38}$, 
C.~Abellan~Beteta$^{33,n}$, 
A.~Adametz$^{11}$, 
B.~Adeva$^{34}$, 
M.~Adinolfi$^{43}$, 
C.~Adrover$^{6}$, 
A.~Affolder$^{49}$, 
Z.~Ajaltouni$^{5}$, 
J.~Albrecht$^{35}$, 
F.~Alessio$^{35}$
% ...

{\footnotesize \it
$ ^{1}$Centro Brasileiro de Pesquisas F\'{i}sicas (CBPF), Rio de Janeiro, Brazil\\
% ...
$ ^{5}$Clermont Universit\'{e}, Universit\'{e} Blaise Pascal, CNRS/IN2P3, LPC, Clermont-Ferrand, France\\
$ ^{6}$CPPM, Aix-Marseille Universit\'{e}, CNRS/IN2P3, Marseille, France\\
% ...
$ ^{11}$Physikalisches Institut, Ruprecht-Karls-Universit\"{a}t Heidelberg, Heidelberg, Germany\\
% ...
$ ^{33}$Universitat de Barcelona, Barcelona, Spain\\
$ ^{34}$Universidad de Santiago de Compostela, Santiago de Compostela, Spain\\
$ ^{35}$European Organization for Nuclear Research (CERN), Geneva, Switzerland\\
% ...
$ ^{38}$Nikhef National Institute for Subatomic Physics, Amsterdam, The Netherlands\\
% ...
$ ^{43}$H.H. Wills Physics Laboratory, University of Bristol, Bristol, United Kingdom\\
% ...
$ ^{49}$Oliver Lodge Laboratory, University of Liverpool, Liverpool, United Kingdom\\
% ...
$ ^{n}$LIFAELS, La Salle, Universitat Ramon Llull, Barcelona, Spain\\
% ...
}
\end{flushleft}
%%%%%%%%%%%%%%%%%%%%%%%%%%%%%%%%%%%%%%%%%%

\cleardoublepage

\renewcommand{\thefootnote}{\arabic{footnote}}
\setcounter{footnote}{0}

%%%%%%%%%%%%%%%%%%%%%%%%%
%%%%% Main text %%%%%%%%%
%%%%%%%%%%%%%%%%%%%%%%%%%

\pagestyle{plain} % restore page numbers for the main text
\setcounter{page}{1}
\pagenumbering{arabic}

\section{Introduction}
\label{sec:Introduction}
The ratio of fragmentation fractions \fsfdt quantifies the relative production rate of \Bs mesons  with respect to \Bz mesons.
Knowledge of this quantity is essential when determining any \Bs branching fraction at the \lhc.
The measurement of the branching fraction of the rare decay \Bsmm~\cite{Aaij:2012ct} is the prime
example where a precise measurement of \fsfdt is crucial for reaching the highest sensitivity in
the search for physics beyond the Standard Model.  
The branching fractions of a large number of \Bd and $B^+$ decays have been measured to high
precision at the \B factories~\cite{PDG2012}, but no \Bs branching fraction is yet known with 
sufficiently high precision to be used as a normalisation channel.

The relative production rates of $b$ hadrons are determined by the fragmentation fractions
$f_u$, $f_d$, $f_s$, $f_c$ and $f_\L$, which describe the probability that a
$b$ quark will hadronize into a $B_q$ meson (where $q=u,d,s,c$), or a 
$b$ baryon, respectively\footnote{Charge conjugation is implied
throughout this paper.}.
The ratio of fragmentation fractions \fsfdt has been previously measured at LHCb with hadronic~\cite{Aaij:2011hi} 
and semileptonic decays~\cite{Aaij:2011jp}, and the
resulting values were combined~\cite{Aaij:2011jp}. 

In this paper, the ratio of fragmentation fractions \fsfdt is determined using
\BsDp and \BdDK decays collected in $pp$ collisions at a centre-of-mass
energy of $\sqrt{s}=7$~TeV, with data corresponding to an integrated luminosity
of $1.0~\fb^{-1}$ recorded with the LHCb detector.  Since the ratio of branching
fractions of the two decay channels is theoretically well
understood~\cite{Fleischer:2010ay}, their relative decay rates can be used to
determine the ratio of fragmentation fractions for \Bs and \Bz mesons through
\begin{eqnarray}\label{eq:fsfd-theo}
\frac{f_s}{f_d} & = &
\frac{\BR(\BdDK)}{\BR(\BsDp)}
\frac{\epsilon_{D K}}{\epsilon_{D_s \pi}}
 \frac{N_{D_s\pi}} {N_{D K}} \nonumber \\
 & = &
\Phi_{\mathrm{PS}} \left|\frac{V_{us}}{V_{ud}}\right|^2 \left(\frac{f_K}{f_{\pi}}\right)^2 \frac{\tau_{B^0}}{\tau_{B^0_s}}
\frac{1}{{\cal N}_a {\cal N}_F}
\frac{\BR(D^- \rightarrow K^+ \pi^- \pi^-)}{\BR(D_s^- \rightarrow K^+ K^- \pi^-)}
\frac{\epsilon_{D K}}{\epsilon_{D_s \pi}}
 \frac{N_{D_s\pi}} {N_{D K}},
\end{eqnarray} 
where $N$ corresponds to a signal yield, $\epsilon$ corresponds to a total
efficiency, $\tau_{B_s^0}/\tau_{B^0}=0.984 \pm 0.011$~\cite{Amhis:2012bh}
corresponds to the ratio of lifetimes and $\BR(D^-\to K^+\pi^-\pi^-)=(9.14\pm
0.20)\%$~\cite{CleoBRD} and $\BR(D_s^-\to K^+K^-\pi^-)=(5.50\pm
0.27)\%$~\cite{Alexander:2008aa} correspond to the $D^-_{(s)}$ meson branching
fractions.  The factor ${\cal N}_a = 1.00 \pm 0.02$ accounts for the ratio of
non-factorizable corrections~\cite{Fleischer:2010-2}, ${\cal N}_F = 1.092 \pm
0.093$ for the ratio of $B^0_{(s)}\to D^-_{(s)}$ form
factors~\cite{Bailey:2012rr}, and $\Phi_{\mathrm{PS}} = 0.971$ for the
difference in phase space due to the mass differences of the initial and final state
particles.  The numerical values used for the CKM matrix elements are $|V_{us}|
= 0.2252$, $|V_{ud}| = 0.97425$, and for the decay constants are $f_{\pi} =
130.41$\,MeV, $f_K = 156.1$\,MeV, with negligible uncertainties, below 1\%~\cite{PDG2012}.  
The measurement is not statistically limited by the size of the
\BdDK sample , and therefore the theoretically
less clean \BdDp decays, where exchange diagrams contribute to the total
amplitude, do not contribute to the knowledge of \fsfdt.

The ratio of fragmentation fractions can depend on the centre-of-mass energy,
as well as on the kinematics of the $B^0_{(s)}$ meson,
as was studied previously at LHCb with partially
reconstructed $B$ decays~\cite{Aaij:2011jp}.  The dependence of the ratio of fragmentation
fractions on the transverse momentum \pt and pseudorapidity $\eta$ of the
$B^0_{(s)}$ meson is determined using fully reconstructed \BdDp
and \BsDp decays. Since it is only the dependence that is of interest here, the more
abundant \BdDp decay is used rather than the \BdDK decay.
The \BdDK and \BdDp decays are also used to determine their ratio of
branching fractions, which can be used to quantify non-factorizable effects in
such heavy-to-light decays~\cite{Fleischer:2010-2}.

The paper is organised as follows: the detector is described in Sec.~\ref{sec:Detector}, followed by the
event selection and the relative selection efficiencies in
Sec.~\ref{sec:selection}.  The fit to the mass distributions and the
determination of the signal yields are discussed in Sec.~\ref{sec:fit}. The
systematic uncertainties are presented in Sec.~\ref{sec:sys}, and the final
results are given in Sec.~\ref{sec:result}.

\section{Detector and software}
\label{sec:Detector}

The \lhcb detector~\cite{Alves:2008zz} is a single-arm forward
spectrometer covering the \mbox{pseudorapidity} range $2<\eta <5$,
designed for the study of particles containing \bquark or \cquark
quarks. The detector includes a high precision tracking system
consisting of a silicon-strip vertex detector surrounding the $pp$
interaction region, a large-area silicon-strip detector located
upstream of a dipole magnet with a bending power of about
$4{\rm\,Tm}$, and three stations of silicon-strip detectors and straw
drift tubes placed downstream. Data are taken with both magnet polarities.
The combined tracking system has
momentum resolution $\Delta p/p$ that varies from 0.4\% at 5\gevc to
0.6\% at 100\gevc, and impact parameter\footnote{Impact parameter~(IP) is 
defined as the transverse distance of closest approach between the 
track and a primary interaction.} resolution of 20\mum for
tracks with high transverse momentum. Charged hadrons are identified
using two ring-imaging Cherenkov detectors. 

The trigger~\cite{Aaij:2012me} consists of a hardware stage, based on
information from the calorimeter and muon systems, followed by a software stage
which applies a full event reconstruction.  The events used in this analysis are
selected at the hardware stage by requiring a cluster in the
calorimeters with transverse energy larger than 3.6~GeV.  The software stage
requires a two-, three- or four-track secondary vertex with a high sum of the
\pt of the tracks and a significant displacement from the primary $pp$
interaction vertices~(PVs).  At least one track should have \pt greater than
$1.7\gevc$, track fit $\chisq$ over the number of degrees of freedom less than two,
and IP \chisq with respect to the
associated primary interaction greater than sixteen. The IP \chisq is defined as the difference
between the \chisq from the vertex fit of the associated PV reconstructed with
and without the considered track. A multivariate algorithm is used for the
identification of the secondary vertices consistent with the decay of a \bquark
hadron.

In the simulation, $pp$ collisions are generated using
\pythia~6.4~\cite{Sjostrand:2006za} with a specific \lhcb
configuration~\cite{LHCb-PROC-2010-056}.  Decays of hadronic particles
are described by \evtgen~\cite{Lange:2001uf}, whilst final state
radiation is generated using \photos~\cite{Golonka:2005pn}. The
interaction of the generated particles with the detector and its
response are implemented using the \geant
toolkit~\cite{Allison:2006ve, *Agostinelli:2002hh} as described in
Ref.~\cite{LHCb-PROC-2011-006}.

\section{Event selection}
\label{sec:selection}

The three decay modes, $B^0\to D^- \pi^+$, $B^0\to D^- K^+$ and $B^0_s\to D_s^-
\pi^+$, are topologically very similar and can therefore be selected using the
same event selection criteria, thus minimizing efficiency differences between
the modes.  The $\B^0_{(s)}$ candidates are reconstructed from a $D_{(s)}^-$
candidate and an additional pion or kaon (the ``bachelor'' particle), with the
$D_{(s)}^-$ meson decaying to $K^+ \pi^- \pi^-$ ($K^+ K^- \pi^-$).

After the trigger selection, a loose preselection is made using the $B^0_{(s)}$
and $D_{(s)}^-$ masses, lifetimes and vertex qualities.  A 
boosted decision tree~(BDT)~\cite{Breiman} is used to further separate
signal from background.  The BDT is trained on half the \BsDp data sample. The
most discriminating variables are the $B^0_{(s)}$ impact parameter \chisq, 
the pointing angle of the $B^0_{(s)}$ candidate to the primary vertex and the
\pt of the tracks.  A cut value for the BDT output variable was chosen to
optimally reduce the number of combinatorial background events, retaining
approximately 84\% of the signal events.

The $D_{(s)}^-$ candidates are identified by requiring the invariant mass under
the $K^+\pi^-\pi^-$ ($K^+K^-\pi^-$) hypothesis to fall within the selection
window 1844 -- 1890 (1944 -- 1990)~\mevcc.  The relative efficiency of the
selection procedure is evaluated for all decay modes using simulated events,
generated with the appropriate Dalitz
plot structures~\cite{delAmoSanchez:2010yp,Bonvicini:2008jw}.  Since the analysis is only
sensitive to relative efficiencies, the impact of any discrepancy between data
and simulation is small.

The final \BsDp and \BdDp event samples are obtained after a particle
identification~(PID) criterion, based on the difference
in log-likelihood between the kaon and pion hypotheses (DLL).
A cut on the bachelor particle is placed at DLL($K-\pi$)$<0$. 
The \mbox{\BdDK} sample is selected by requiring DLL($K-\pi$)$>5$ for
the bachelor particle.  The \mbox{$D_s^-\to K^+K^-\pi^-$} decay is distinguished from
$D^-\to K^+\pi^-\pi^-$ decays by imposing DLL($K-\pi$)$>5$ on the kaon candidate
with the same charge as the $D$ meson, whilst the DLL criteria for the
$\pi^{-}$ and $K^+$ are identical between $D^-$ and $D_s^-$ and
are used to discriminate $D^-_{(s)}$ decays from background.  The PID
performance as a function of \pt and $\eta$ of the track is estimated from data
using a calibration sample of approximately 27 million $D^{*-}\to
\Dzb(K^+\pi^-)\pi^-$ decays, which are selected using kinematic criteria only.

\section{Event yields}
\label{sec:fit}

\begin{figure}[!t]
 \centering
 \begin{picture}(250,480)(0,0) 
   \put(-30,336){\includegraphics[width=.80\textwidth]{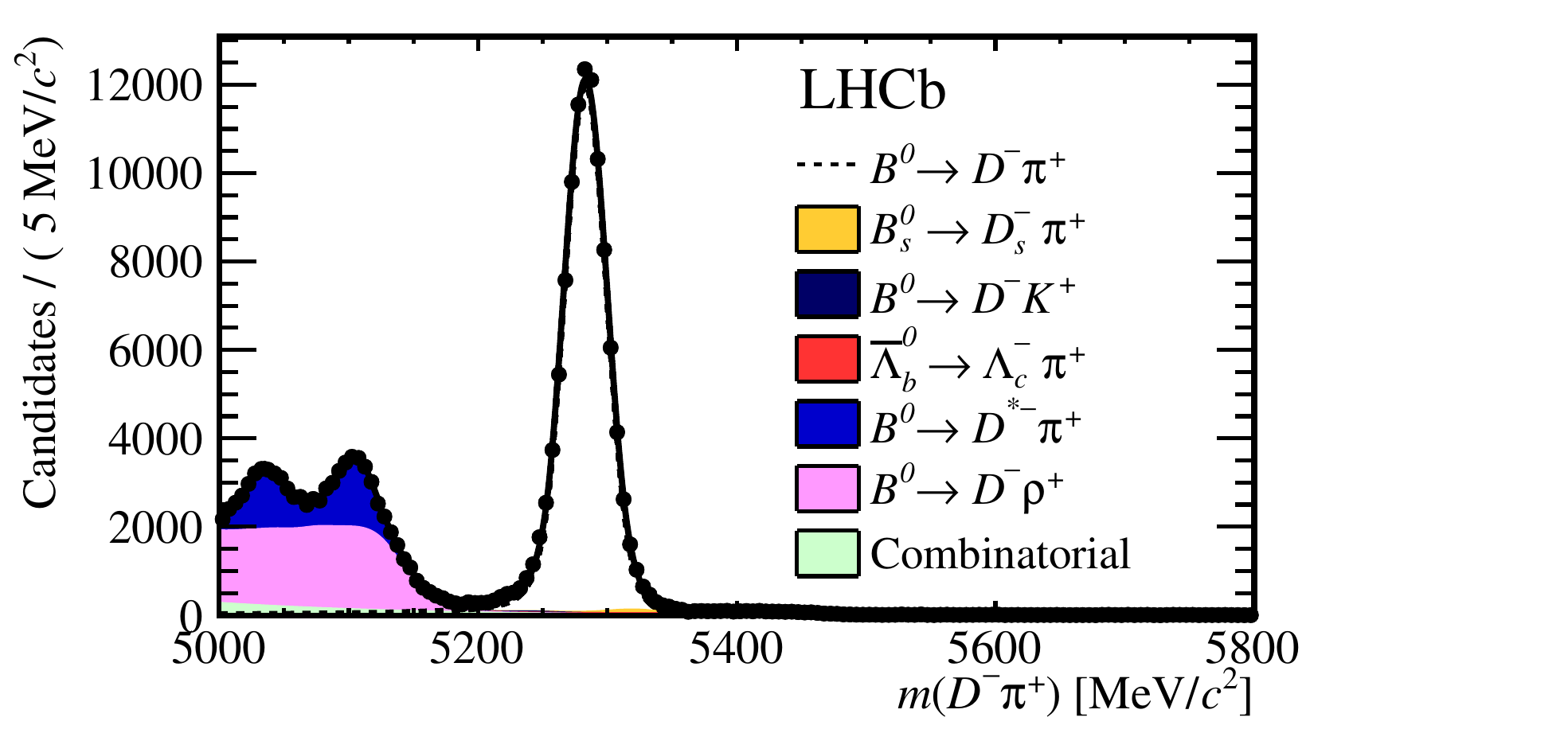}}
   \put(-30,168){\includegraphics[width=.80\textwidth]{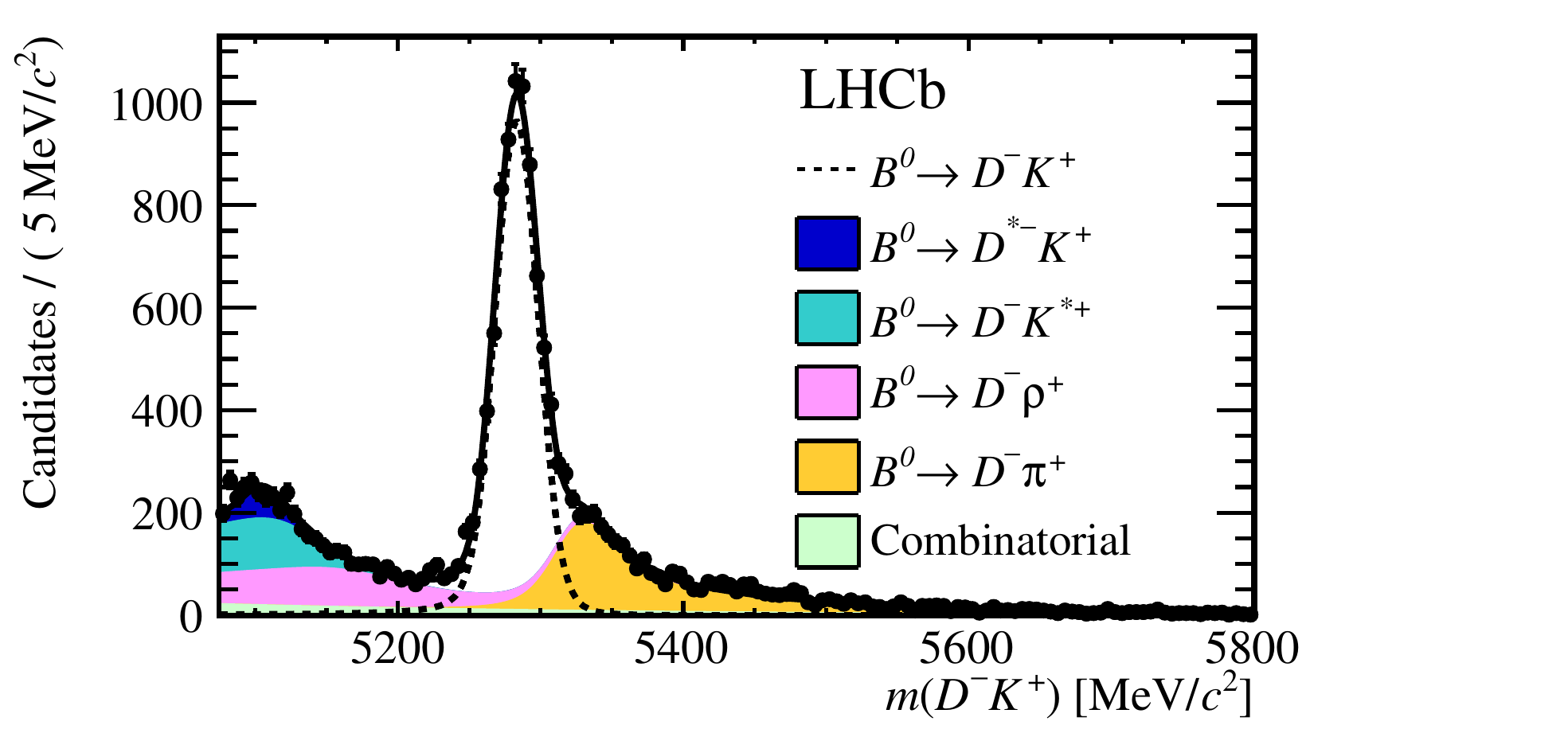}}
   \put(-30,0){\includegraphics[width=.80\textwidth]{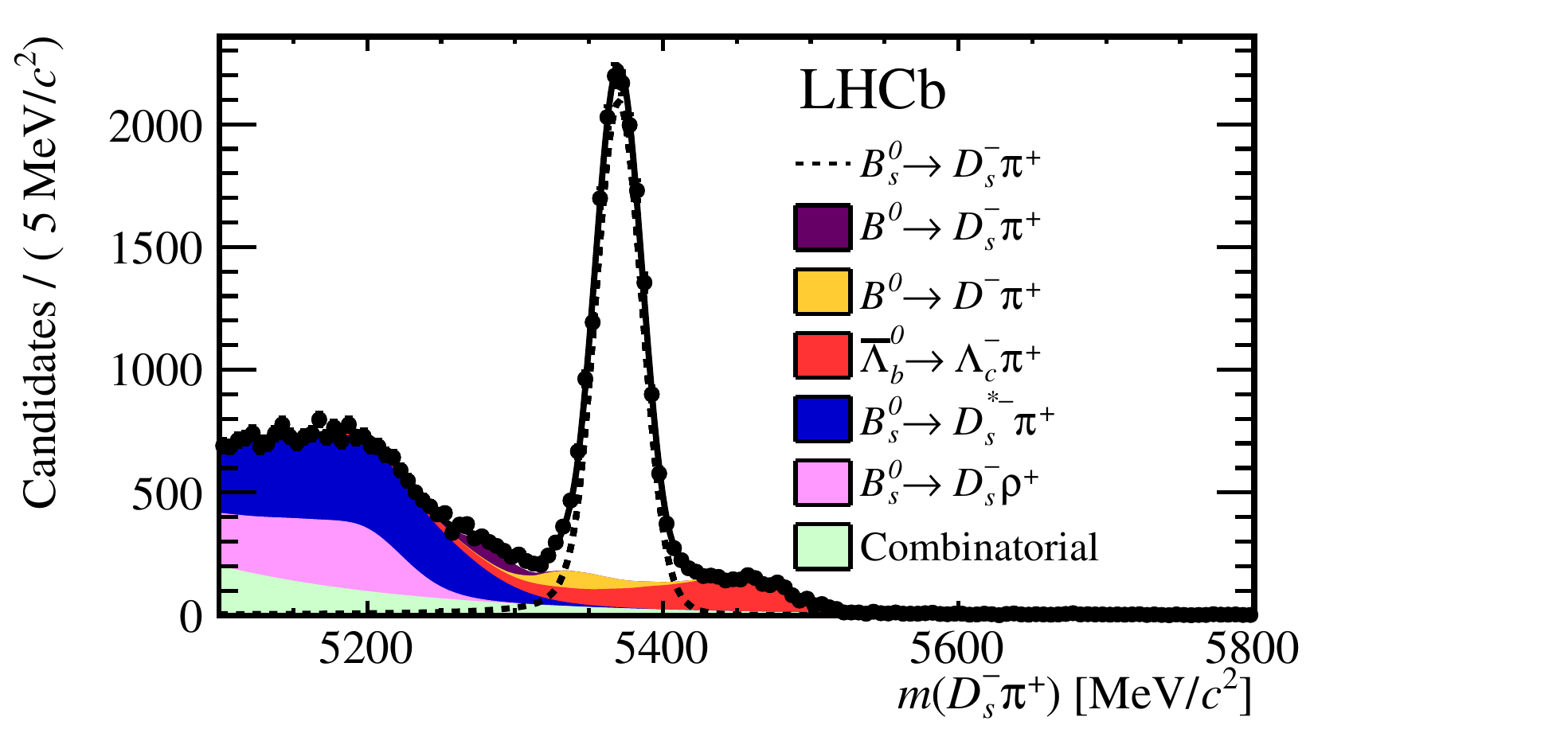}}
   \put(40,470){(a)}
   \put(40,304){(b)}
   \put(40, 137){(c)}
 \end{picture} 
 \caption{
   Invariant mass distributions of (a) \BdDp
   (b) \BdDK and (c) \BsDp candidates. The solid line is the result of the fit and the dotted
   line indicates the signal. 
   The stacked background shapes follow the same top-to-bottom order in the legend and the plot.
   The \Bs and \Lbbar backgrounds in the \BdDp mass distribution are invisibly small.
   The resulting signal yields are listed in Table~\ref{tab:yields}.
   For illustration purposes the figures include events from both magnet polarities, although they are fitted
   separately as described in the text.}
 \label{fig:fits}
\end{figure}

The relative yields of the three decay modes are determined from unbinned
extended maximum likelihood fits to the mass distributions of the reconstructed
$B^0_{(s)}$ candidates as shown in Fig.~\ref{fig:fits}.  In order to achieve the
highest sensitivity, the sample is separated according to the two magnet
polarities, allowing for possible differences in PID performance and in running
conditions. A simultaneous fit to the two magnet polarities is performed for
each decay mode, with the peak position and width of each signal shared between
the two.

The signal mass shape is described by a Gaussian distribution with power-law
tails on either side to model the radiative tail and non-Gaussian detector
effects. It consists of a Crystal Ball function~\cite{Skwarnicki:1986xj}

\begin{eqnarray}
f_\mathrm{left}(m,\alpha,n,\mu,\sigma) = N \cdot \left\{ 
 \begin{array}{ll}
\label{eq:DCB}
   e^{-\frac{(m-\mu)^2}{2\sigma^2}},
            & \mathrm{for}\;  \frac{m-\mu}{\sigma}>-\alpha\\
   \left(\frac{n}{|\alpha|} \right)^{n} \cdot e^{-|\alpha|^2/2} \cdot 
   \left(\frac{n}{|\alpha|} - |\alpha| - \frac{m-\mu}{\sigma}\right)^{-n},
            & \mathrm{for}\;  \frac{m-\mu}{\sigma}\leq -\alpha \\
 \end{array}
\right.  \nonumber\\
\end{eqnarray}
and a second, similar but mirrored, function to describe
the right tail, resulting in the signal mass shape
$f_\mathrm{2CB}(m)=f_\mathrm{left}(m)+f_\mathrm{right}(m)$.  The parameters of
the tails are obtained from simulated events. The mean $\mu$ and the width
$\sigma$ of the Gaussian distribution are equal in both Crystal Ball functions,
and are allowed to vary in the fit. The parameter $N$ is a normalisation factor.

Three classes of background are considered in the fit: fully reconstructed
decays where at least one track is misidentified, partially reconstructed decays
with or without misidentified tracks and combinatorial background.  The shapes
of the invariant mass distributions for the partially reconstructed decays are
taken from large samples of simulated events.  The main sources are \BdDrho and
$\decay{\Bz}{\Dstarm \pip(\Kp)}$ for the $\decay{\Bz}{\Dm \pip(\Kp)}$ sample,
and \BsDrho and \BsDstarp for the \BsDp sample.

The invariant mass distributions of the misidentified decays are affected by the
PID criteria. The shapes are obtained from simulated events, with the
appropriate mass hypothesis applied. The distribution is then reweighted in a
data-driven way, according to the particle identification cut efficiency
obtained from the calibration sample, which is strongly dependent on the
momentum of the particle.

Despite the small $\pi \to K$ misidentification probability of 2.8\%, the
largest misidentified background in the \BdDK sample originates from Cabibbo-favoured \BdDp
decays where the bachelor pion is misidentified as a kaon.  
The shape of this
particular misidentified decay is determined from data using a high purity
sample of \BdDp decays (see Fig.~\ref{fig:fits}(a)), 
obtained by selecting events in a narrow mass window 5200--5340~\mevcc.  
The yield of this prominent peaking
background is allowed to vary in the fit and is found to be consistent with the
expected yield based on the \BdDp signal yield and the misidentification
probability.  The contamination of \BdDp events in the \BsDp sample can be
caused by the misidentification of either pion from the \Dm decay. The
misidentification probability is 2.0\% (3.2\%) for the higher (lower) \pt
pion.  
After selecting the \Dsm candidate within the mass window around the nominal 
\Dsm mass~\cite{PDG2012},
the number of misidentified pions is reduced to 0.75\% (0.02\%). 
The yield of this background is constrained in the fit, based on the \BdDp
signal yield, the misidentification probability and their associated
uncertainties.

The yield of \LbbarLcp decays is allowed to vary in the fit.  The cross-feeds from
\BdDK and \BsDp events in the \BdDp signal is small, and are constrained to their
respective predicted yields.  In addition, a contribution from the rare \BdDsp decay
is expected with a yield of 3.3\% compared to the \BsDp signal, and is accounted for
accordingly.

\begin{table}[!t]
  \begin{center}
  \label{tab:yields}   
  \caption{Yields obtained from the fits to the invariant mass distributions.}
  \begin{tabular}{ lr }
    Signal  & Yield                     \\
    \hline    
    \BdDp   & $106\,197\pm 344 $        \\
    \BdDK   & $  7\,664\pm \,\,\,  99 $ \\
    \BsDp   & $ 17\,419\pm 155 $        \\
    \end{tabular}
  \end{center}
\end{table}

The combinatorial background consists of events with random pions and kaons,
forming a fake \Dm or \Dsm candidate, as well as real \Dm or \Dsm mesons, that
combine with a random pion or kaon. The combinatorial background is modelled
with an exponential shape.

The results of the fits are presented in Fig.~\ref{fig:fits}, and the
corresponding signal yields are listed in Table~\ref{tab:yields}. The total
yields of the decays \BdDp and \BdDK are used to determine the ratio of their
branching fractions, while the event yields of the
decays \BsDp and \BdDK are used to measure the average ratio of fragmentation
fractions.

The dependence of the relative $b$-hadron production fractions as a function of
the transverse momentum and pseudorapidity of the $B^0_{(s)}$ meson is studied
in the ranges $2.0 < \eta < 5.0$ and $1.5 < \pt < 40$ \gevc, using \BdDp and
\BsDp decays.  The event sample is subdivided in 20 bins in \pt and 10 bins in
$\eta$, with the bin sizes chosen to obtain approximately equal number of events
per bin.  The fitting model for each bin is the same as that for the
integrated samples, apart from the treatment of the exponent of the
combinatorial background distribution, which is fixed to the value obtained from
the fits to the integrated sample.

\section{Systematic uncertainties}
\label{sec:sys}

The systematic uncertainties on the measurement of the relative event yields
of the \BdDp, \BdDK and \BsDp decay modes are related to trigger and offline selection efficiency 
corrections, particle identification calibration and the fit model.

\begin{table}[!b]
  \begin{center}
  \caption{Systematic uncertainties for the measurement of the corrected 
    ratio of event yields used for the measurements of \fsfdt and the relative
    branching fraction of \BdDK. The systematic uncertainty in \pt and $\eta$ bins is shown as a range in 
    the last column, and the total systematic uncertainty is 
    the quadratic sum of the uncorrelated uncertainties. 
    The systematic uncertainties on the ratio of \BdDp and \BsDp yields that are correlated 
    among the bins do not affect the dependence on \pt or $\eta$, and are not accounted for in
    the total systematic uncertainty.}
  \label{t:sysbudgettotal}   
    \begin{tabular}{ lccc }
    Source  \TVA \BVA        & $\frac{\BdDp}{\BdDK}$ (\%) & $\frac{\BsDp}{\BdDK}$ (\%) &$\frac{\BsDp}{\BdDp}$ (\%)    \\
    \hline
    Detector acceptance    &                      &                        &                     \\
\hspace{0.5cm} and reconstruction & $0.7$         & $0.7$                  & $2.0-2.9$           \\
    Hardware trigger efficiency   & $2.0$         & $2.0$                  & $0.8$               \\
    Offline selection      & $1.2$                & $1.1$                  & $1.2$               \\
    BDT cut                & $1.0$                & $1.0$                  & $1.5$               \\
    PID selection          & $1.0$                & $1.5$                  & $1.1$               \\
    Comb. background       & $0.7$                & $1.0$                  & $0.8$               \\
    Signal shape (tails)   & $0.5$                & $0.6$                  & [correl.]           \\
    Signal shape (core)    & $0.8$                & $1.0$                  & [correl.]           \\
    Charmless background   & $0.4$                & --                     & [correl.]           \\
    \hline
    Total                  & $3.1$                & $3.4$                  & $3.2-3.8$           \\
    \end{tabular}
  \end{center}
\end{table}

The response to charged pions and kaons of the hadronic calorimeter used at the
hardware trigger level has been investigated. As the hardware trigger mostly
triggers on the high-\pt bachelor, a systematic uncertainty of 2\% is assigned
to the ratio of trigger efficiencies for the decays \BdDK and \BdDp, estimated
from dedicated studies with $D^{*-}\to \Dzb(K^+\pi^-)\pi^-$ decays. This
uncertainty is assumed to be uncorrelated between the individual bins in the binned analysis.

The relative selection efficiencies from simulation are studied by varying the
BDT criterion, changing the signal yields by about $\pm 25\%$. The variation of
the relative efficiency is 1.0\% which is assigned as systematic uncertainty.

The uncertainty on the PID efficiencies is estimated by comparing, in simulated
events, the results obtained using the $D^{*-}$ calibration sample to the true
simulated PID performance on the signal decays. The corresponding uncertainty
ranges from 1.0\% to 1.5\% for the different measurements.

The exponent of the combinatorial background distribution is allowed to vary in
the fits to the \BdDp and \BsDp mass distributions. 
By studying $\Dm\pi^-$ and $\Dm K^-$ combinations, it is
suggested that the value of the exponent is smaller for the \BdDK decays than for the \BdDp decays,
and therefore in the fit to the \BdDK candidates the exponent is fixed to half 
the value found in the fit to the \BdDp sample.
The uncertainty on the signal yields due to the shape of the combinatorial
background is estimated by reducing the exponent to half its value in the fits to
the \BdDp and \BsDp mass distributions, and by taking a flat background for the
fit to the \BdDK mass distribution. An uncertainty of 1.0\% (0.7\%) is assigned
to the relative \BdDK and \BsDp (\BdDp) yields.

The tails of the signal distributions are fixed from simulation due to the
presence of large amounts of partially reconstructed decays in the lower
sidebands. The uncertainty on the signal yield is estimated by varying the
parameters that describe the tails by 10\%.  The uncertainty from the shape of
the central peak is taken from a fit allowing for two different widths for the
Crystal Ball functions in Eq.~\ref{eq:DCB}, leading to a 1.0\% (0.8\%) uncertainty
on the relative \BdDK and \BsDp (\BdDp) yields.

The contribution of \emph{charmless} $B$ decays without an intermediate \D meson
is ignored in the fit.  To evaluate the systematic uncertainty due to these
decays, the \B mass spectra for candidates in the sidebands of the $D$ mass
distribution are examined. A contribution of 0.4\% relative to the signal yield
is found in the \BdDp decay mode, and no contribution is seen in the other
modes.  For the \BdDp decay mode no correction is applied and the full size is
taken as an uncertainty. No systematic uncertainty is assigned for the other
decay modes.

The various sources of the systematic uncertainty that contribute to the
uncertainties on the ratios of signal yields are listed in
Table~\ref{t:sysbudgettotal}.  No uncertainty is associated to the \LbbarLcp 
background, as the yield is allowed to vary in the
fit. Other cross checks, like varying the \BdDsp yield in the \BsDp fit or
including \LbbarLcK in the \BdDK fit, show a negligible effect on the signal
yields.

All systematic variations are also performed in bins, and the corresponding
relative changes in the ratio of yields have been quantified.  Variations
showing correlated behaviour do not affect the slope and are therefore not
considered further.

\section{Results}
\label{sec:result}
The relative signal yields of the decays \BdDp, \BdDK and \BsDp are used to
determine the branching fraction of the decay \BdDK, and the ratio of
fragmentation fractions \fsfdt.

The efficiency corrected ratio of \BdDK and \BdDp signal yields results in the
ratio of branching fractions
\begin{equation}\nonumber
\frac{\BR\left(\BdDK\right)}{\BR\left(\BdDp\right)} = 0.0822 \pm 0.0011 \, (\textrm{stat}) \pm 0.0025 \, (\textrm{syst}).
\end{equation}
This is combined with the world average branching fraction
$\BR\left(\BdDp\right) = (26.8 \pm 1.3)\times 10^{-4}$~\cite{PDG2012}, to give
\begin{equation}
\BR\left(\BdDK\right) = (2.20 \pm 0.03  \pm 0.07  \pm 0.11)\times 10^{-4},
\nonumber
\end{equation}
where the first uncertainty is statistical, the second is systematic and the last is
due to the uncertainty on the \BdDp branching fraction.

The ratio of fragmentation fractions is determined from the efficiency corrected
event yields. The ratio of efficiencies is $0.913\pm 0.027$.  This results in
\begin{eqnarray} 
\label{eq:fsfdt_NF_DK}
\nonumber
\fsfd & = & (0.261 \pm 0.004  \pm 0.017 ) \times  \frac{1}{{\cal N}_a {\cal N}_F} \\
      & = &  0.238 \pm 0.004  \pm 0.015 \pm 0.021 \, ,
\nonumber
\end{eqnarray}
where the first uncertainty is statistical, the second is systematic containing
the sources listed in Table \ref{t:sysbudgettotal} as well as errors from external
measurements, and the third is theoretical, due to the knowledge of ${\cal
N}_a$ and ${\cal N}_F$.  The last source is dominated by the uncertainty on the form
factor ratio.

This measurement supersedes and is in agreement with the previous determination
with hadronic decays~\cite{Aaij:2011hi}.  It also agrees with the previous
measurement based on semileptonic decays~\cite{Aaij:2011jp}.  The 
two independent results are combined taking into account the various
sources of correlated systematic uncertainties, notably the $D^-_{(s)}$ branching
fractions and $B^0_{(s)}$ lifetimes, to give
\begin{equation}
\label{eq:fsfd_com}
\fsfd = 0.256 \pm 0.020,
\end{equation}
which supersedes the previous measurement from LHCb.

\begin{figure}[!t]
 \centering
 \begin{picture}(250,150)(0,0) 
   \put(-105,0){\includegraphics[width=.49\textwidth]{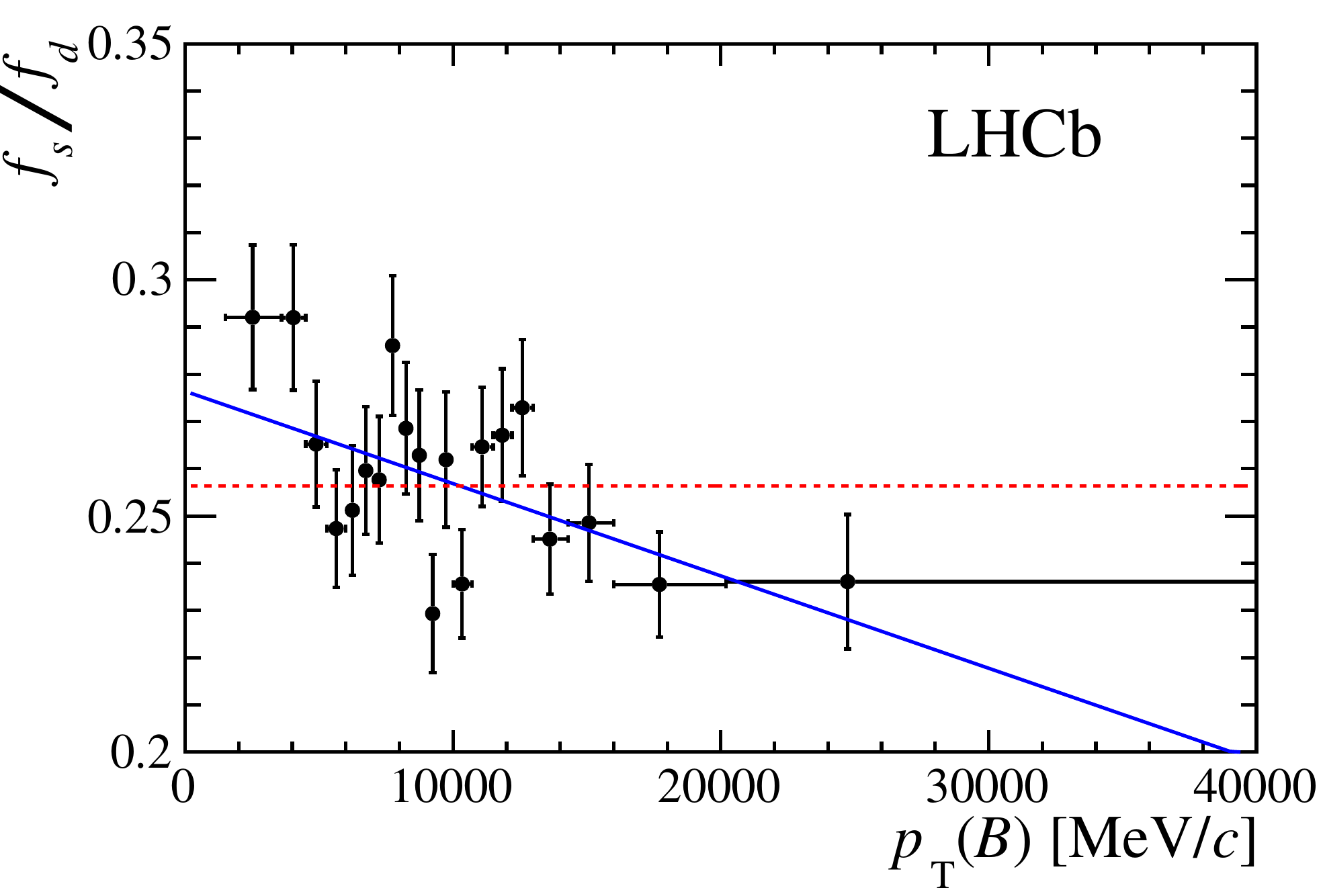}}
   \put(130,0){\includegraphics[width=.49\textwidth]{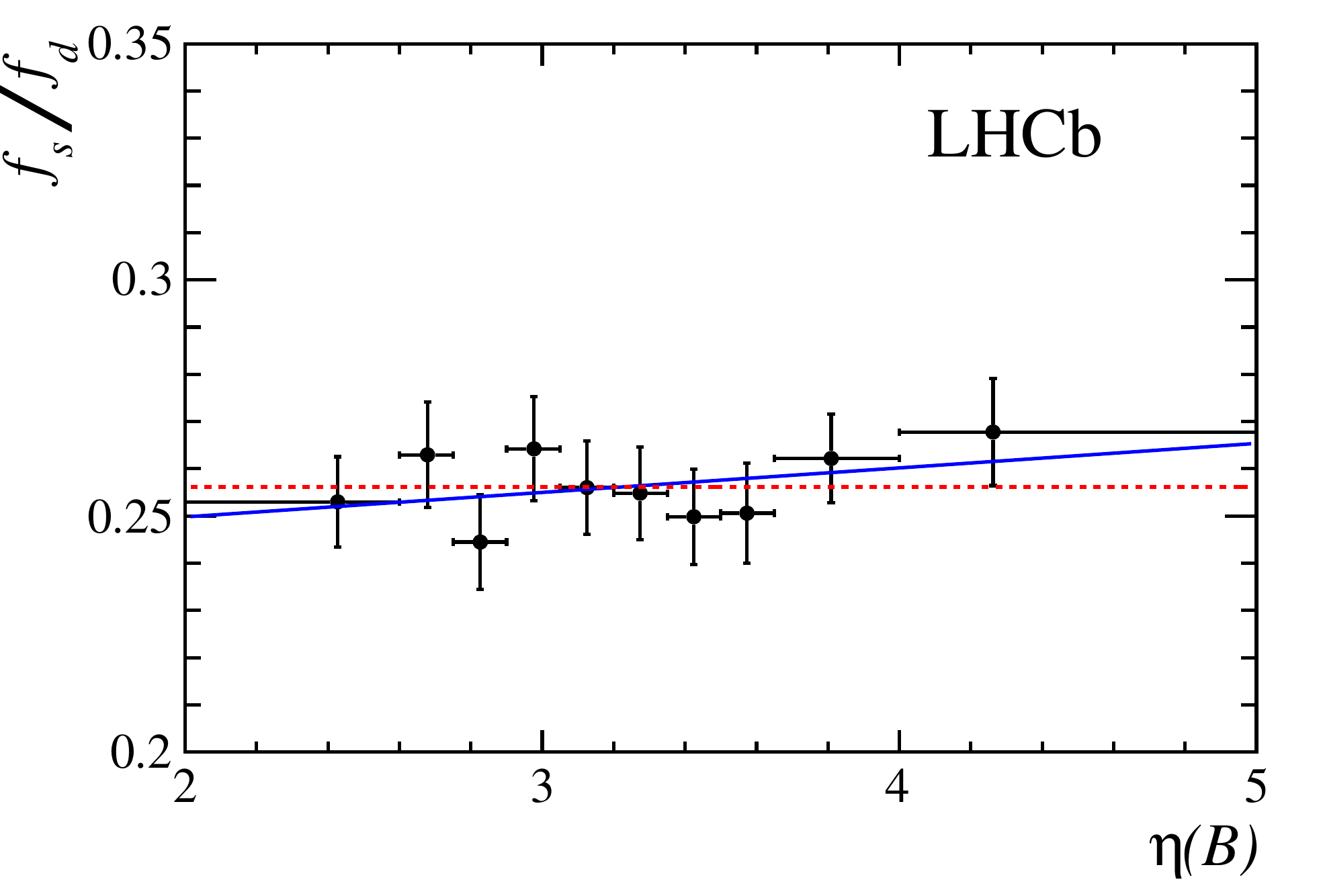}}
   \put(-65,125){(a)}
   \put(170,125){(b)}
 \end{picture}
 \caption{Ratio of fragmentation fractions \fsfdt as functions of (a) \pt and (b) $\eta$.
 The errors on the data points are the statistical and uncorrelated systematic uncertainties added in quadrature.
 The solid line is the result of a linear fit, and 
 the dashed line corresponds to the fit for the no-dependence hypothesis.  The average value of \pt or $\eta$ 
 is determined for each bin and used as the center of the bin. The horizontal error bars indicate the bin size. Note 
 that the scale is zero suppressed. 
}
 \label{fig:fsfd_dep}
\end{figure}

The value of \fsfdt in bins of \pt or $\eta$ is determined using the \BsDp and \BdDp decay modes 
and is presented in Fig.~\ref{fig:fsfd_dep}. A linear $\chi^2$ fit gives

\begin{eqnarray}\label{eq:fsfd-dep}
\fsfdt(\pt)  
             & = & (0.256\pm 0.020) + (-2.0\pm 0.6)\times 10^{-3}/\gev/c \times (\pt  - \langle\pt\rangle)\nonumber\\
\fsfdt(\eta) 
             & = & (0.256\pm 0.020) + (0.005\pm 0.006)\times  (\eta  - \langle\eta\rangle), \nonumber
\end{eqnarray} 
with $\langle\pt\rangle=10.4$~GeV/$c$ and $\langle\eta\rangle =3.28$.  The data points
are normalised with a scale factor to match the average value of 0.256. The
uncertainty associated to this parameter is taken from Eq. \ref{eq:fsfd_com},
whilst the error from the fit is 0.003 for both \pt and $\eta$.

The p-value for this linear fit is found to be 0.16 (0.87) for \pt ($\eta$).
The observed slope for the dependence on the transverse momentum of the $B^0_{(s)}$ meson
deviates from zero with a significance of three standard deviations.  No
indication of a dependence on $\eta(B)$ is found.

\section{Conclusions}
The relative production rate of \Bs and \Bd mesons is determined using the
hadronic decays \BsDp and \BdDK resulting in $\fsfdt = 0.238 \pm
0.004(\textrm{stat}) \pm 0.015(\textrm{syst}) \pm 0.021(\textrm{theo})$.  This
value is consistent with a previous LHCb measurement based on semileptonic decays, 
with which it is averaged to obtain
$\fsfdt = 0.256 \pm 0.020$.  The ratio of fragmentation fractions \fsfdt is
determined as a function of the transverse momentum and pseudorapidity of the
$B^0_{(s)}$ meson, 
and a variation consistent with a linear dependence on the transverse momentum
of the the $B^0_{(s)}$ meson is observed,
with a significance of three standard
deviations.
In addition, the ratio of branching fractions of the decays \BdDK and \BdDp is
measured to be $0.0822 \pm 0.0011 \, (\textrm{stat}) \pm 0.0025 \,
(\textrm{syst})$.

\section*{Acknowledgements}

\noindent We express our gratitude to our colleagues in the CERN
accelerator departments for the excellent performance of the LHC. We
thank the technical and administrative staff at the LHCb
institutes. We acknowledge support from CERN and from the national
agencies: CAPES, CNPq, FAPERJ and FINEP (Brazil); NSFC (China);
CNRS/IN2P3 and Region Auvergne (France); BMBF, DFG, HGF and MPG
(Germany); SFI (Ireland); INFN (Italy); FOM and NWO (The Netherlands);
SCSR (Poland); ANCS/IFA (Romania); MinES, Rosatom, RFBR and NRC
``Kurchatov Institute'' (Russia); MinECo, XuntaGal and GENCAT (Spain);
SNSF and SER (Switzerland); NAS Ukraine (Ukraine); STFC (United
Kingdom); NSF (USA). We also acknowledge the support received from the
ERC under FP7. The Tier1 computing centres are supported by IN2P3
(France), KIT and BMBF (Germany), INFN (Italy), NWO and SURF (The
Netherlands), PIC (Spain), GridPP (United Kingdom). We are thankful
for the computing resources put at our disposal by Yandex LLC
(Russia), as well as to the communities behind the multiple open
source software packages that we depend on.

\addcontentsline{toc}{section}{References}
\bibliographystyle{LHCb}
\bibliography{main-fdfs}

\ifx\mcitethebibliography\mciteundefinedmacro
\PackageError{LHCb.bst}{mciteplus.sty has not been loaded}
{This bibstyle requires the use of the mciteplus package.}\fi
\providecommand{\href}[2]{#2}
\begin{mcitethebibliography}{10}
\mciteSetBstSublistMode{n}
\mciteSetBstMaxWidthForm{subitem}{\alph{mcitesubitemcount})}
\mciteSetBstSublistLabelBeginEnd{\mcitemaxwidthsubitemform\space}
{\relax}{\relax}

\bibitem{Aaij:2012ct}
LHCb collaboration, R.~Aaij {\em et~al.},
  \ifthenelse{\boolean{articletitles}}{{\it {First evidence for the decay
  $B^0_s \to \mu^+ \mu^-$}}, }{}Phys.\ Rev.\ Lett.\  {\bf 110} (2012) 021801,
  \href{http://arxiv.org/abs/1211.2674}{{\tt arXiv:1211.2674}}\relax
\mciteBstWouldAddEndPuncttrue
\mciteSetBstMidEndSepPunct{\mcitedefaultmidpunct}
{\mcitedefaultendpunct}{\mcitedefaultseppunct}\relax
\EndOfBibitem
\bibitem{PDG2012}
Particle Data Group, J.~Beringer {\em et~al.},
  \ifthenelse{\boolean{articletitles}}{{\it {\href{http://pdg.lbl.gov/}{Review
  of particle physics}}},
  }{}\href{http://dx.doi.org/10.1103/PhysRevD.86.010001}{Phys.\ Rev.\  {\bf
  D86} (2012) 010001}\relax
\mciteBstWouldAddEndPuncttrue
\mciteSetBstMidEndSepPunct{\mcitedefaultmidpunct}
{\mcitedefaultendpunct}{\mcitedefaultseppunct}\relax
\EndOfBibitem
\bibitem{Aaij:2011hi}
LHCb collaboration, R.~Aaij {\em et~al.},
  \ifthenelse{\boolean{articletitles}}{{\it {Determination of $f_s/f_d$ for 7
  TeV pp collisions and measurement of the $B^0\to D^-K^+$ branching
  fraction}}, }{}\href{http://dx.doi.org/10.1103/PhysRevLett.107.211801}{Phys.\
  Rev.\ Lett.\  {\bf 107} (2011) 211801},
  \href{http://arxiv.org/abs/1106.4435}{{\tt arXiv:1106.4435}}\relax
\mciteBstWouldAddEndPuncttrue
\mciteSetBstMidEndSepPunct{\mcitedefaultmidpunct}
{\mcitedefaultendpunct}{\mcitedefaultseppunct}\relax
\EndOfBibitem
\bibitem{Aaij:2011jp}
LHCb collaboration, R.~Aaij {\em et~al.},
  \ifthenelse{\boolean{articletitles}}{{\it {Measurement of b hadron production
  fractions in 7 TeV pp collisions}},
  }{}\href{http://dx.doi.org/10.1103/PhysRevD.85.032008}{Phys.\ Rev.\  {\bf
  D85} (2012) 032008}, \href{http://arxiv.org/abs/1111.2357}{{\tt
  arXiv:1111.2357}}\relax
\mciteBstWouldAddEndPuncttrue
\mciteSetBstMidEndSepPunct{\mcitedefaultmidpunct}
{\mcitedefaultendpunct}{\mcitedefaultseppunct}\relax
\EndOfBibitem
\bibitem{Fleischer:2010ay}
R.~Fleischer, N.~Serra, and N.~Tuning,
  \ifthenelse{\boolean{articletitles}}{{\it {New strategy for $B_s$ branching
  ratio measurements and the search for new physics in $B^0_s \to \mu^+
  \mu^-$}}, }{}\href{http://dx.doi.org/10.1103/PhysRevD.82.034038}{Phys.\ Rev.\
   {\bf D82} (2010) 034038}, \href{http://arxiv.org/abs/1004.3982}{{\tt
  arXiv:1004.3982}}\relax
\mciteBstWouldAddEndPuncttrue
\mciteSetBstMidEndSepPunct{\mcitedefaultmidpunct}
{\mcitedefaultendpunct}{\mcitedefaultseppunct}\relax
\EndOfBibitem
\bibitem{Amhis:2012bh}
Heavy Flavor Averaging Group, Y.~Amhis {\em et~al.},
  \ifthenelse{\boolean{articletitles}}{{\it {Averages of b-hadron, c-hadron,
  and $\tau$-lepton properties as of early 2012}},
  }{}\href{http://arxiv.org/abs/1207.1158}{{\tt arXiv:1207.1158}}\relax
\mciteBstWouldAddEndPuncttrue
\mciteSetBstMidEndSepPunct{\mcitedefaultmidpunct}
{\mcitedefaultendpunct}{\mcitedefaultseppunct}\relax
\EndOfBibitem
\bibitem{CleoBRD}
CLEO collaboration, S.~Dobbs {\em et~al.},
  \ifthenelse{\boolean{articletitles}}{{\it {Measurement of absolute hadronic
  branching fractions of $D$ mesons and $e^+e^-\to D\bar{D}$ cross sections at
  the $\psi(3770)$}},
  }{}\href{http://dx.doi.org/10.1103/PhysRevD.76.112001}{Phys.\ Rev.\  {\bf
  D76} (2007) 112001}, \href{http://arxiv.org/abs/0709.3783}{{\tt
  arXiv:0709.3783}}\relax
\mciteBstWouldAddEndPuncttrue
\mciteSetBstMidEndSepPunct{\mcitedefaultmidpunct}
{\mcitedefaultendpunct}{\mcitedefaultseppunct}\relax
\EndOfBibitem
\bibitem{Alexander:2008aa}
CLEO collaboration, J.~Alexander {\em et~al.},
  \ifthenelse{\boolean{articletitles}}{{\it {Absolute measurement of hadronic
  branching fractions of the $D_s^+$ meson}},
  }{}\href{http://dx.doi.org/10.1103/PhysRevLett.100.161804}{Phys.\ Rev.\
  Lett.\  {\bf 100} (2008) 161804}, \href{http://arxiv.org/abs/0801.0680}{{\tt
  arXiv:0801.0680}}\relax
\mciteBstWouldAddEndPuncttrue
\mciteSetBstMidEndSepPunct{\mcitedefaultmidpunct}
{\mcitedefaultendpunct}{\mcitedefaultseppunct}\relax
\EndOfBibitem
\bibitem{Fleischer:2010-2}
R.~Fleischer, N.~Serra, and N.~Tuning,
  \ifthenelse{\boolean{articletitles}}{{\it {Tests of factorization and SU(3)
  relations in $B$ decays into heavy-light final states}},
  }{}\href{http://dx.doi.org/10.1103/PhysRevD.83.014017}{Phys.\ Rev.\  {\bf
  D83} (2011) 014017}, \href{http://arxiv.org/abs/1012.2784}{{\tt
  arXiv:1012.2784}}\relax
\mciteBstWouldAddEndPuncttrue
\mciteSetBstMidEndSepPunct{\mcitedefaultmidpunct}
{\mcitedefaultendpunct}{\mcitedefaultseppunct}\relax
\EndOfBibitem
\bibitem{Bailey:2012rr}
J.~A. Bailey {\em et~al.}, \ifthenelse{\boolean{articletitles}}{{\it {$B_s\to
  D_s/B\to D$ semileptonic form-factor ratios and their application to
  BR($B^0_s\to \mu^+\mu^-$)}},
  }{}\href{http://dx.doi.org/10.1103/PhysRevD.85.114502}{Phys.\ Rev.\  {\bf
  D85} (2012) 114502}, \href{http://arxiv.org/abs/1202.6346}{{\tt
  arXiv:1202.6346}}, Erratum-ibid. {\bf D86} (2012) 039904\relax
\mciteBstWouldAddEndPuncttrue
\mciteSetBstMidEndSepPunct{\mcitedefaultmidpunct}
{\mcitedefaultendpunct}{\mcitedefaultseppunct}\relax
\EndOfBibitem
\bibitem{Alves:2008zz}
LHCb collaboration, A.~A. Alves~Jr. {\em et~al.},
  \ifthenelse{\boolean{articletitles}}{{\it {The \lhcb detector at the LHC}},
  }{}\href{http://dx.doi.org/10.1088/1748-0221/3/08/S08005}{JINST {\bf 3}
  (2008) S08005}\relax
\mciteBstWouldAddEndPuncttrue
\mciteSetBstMidEndSepPunct{\mcitedefaultmidpunct}
{\mcitedefaultendpunct}{\mcitedefaultseppunct}\relax
\EndOfBibitem
\bibitem{Aaij:2012me}
R.~Aaij {\em et~al.}, \ifthenelse{\boolean{articletitles}}{{\it {The \lhcb
  trigger and its performance}}, }{}\href{http://arxiv.org/abs/1211.3055}{{\tt
  arXiv:1211.3055}}\relax
\mciteBstWouldAddEndPuncttrue
\mciteSetBstMidEndSepPunct{\mcitedefaultmidpunct}
{\mcitedefaultendpunct}{\mcitedefaultseppunct}\relax
\EndOfBibitem
\bibitem{Sjostrand:2006za}
T.~Sj\"{o}strand, S.~Mrenna, and P.~Skands,
  \ifthenelse{\boolean{articletitles}}{{\it {PYTHIA 6.4 physics and manual}},
  }{}\href{http://dx.doi.org/10.1088/1126-6708/2006/05/026}{JHEP {\bf 05}
  (2006) 026}, \href{http://arxiv.org/abs/hep-ph/0603175}{{\tt
  arXiv:hep-ph/0603175}}\relax
\mciteBstWouldAddEndPuncttrue
\mciteSetBstMidEndSepPunct{\mcitedefaultmidpunct}
{\mcitedefaultendpunct}{\mcitedefaultseppunct}\relax
\EndOfBibitem
\bibitem{LHCb-PROC-2010-056}
I.~Belyaev {\em et~al.}, \ifthenelse{\boolean{articletitles}}{{\it {Handling of
  the generation of primary events in \gauss, the \lhcb simulation framework}},
  }{}\href{http://dx.doi.org/10.1109/NSSMIC.2010.5873949}{Nuclear Science
  Symposium Conference Record (NSS/MIC) {\bf IEEE} (2010) 1155}\relax
\mciteBstWouldAddEndPuncttrue
\mciteSetBstMidEndSepPunct{\mcitedefaultmidpunct}
{\mcitedefaultendpunct}{\mcitedefaultseppunct}\relax
\EndOfBibitem
\bibitem{Lange:2001uf}
D.~J. Lange, \ifthenelse{\boolean{articletitles}}{{\it {The EvtGen particle
  decay simulation package}},
  }{}\href{http://dx.doi.org/10.1016/S0168-9002(01)00089-4}{Nucl.\ Instrum.\
  Meth.\  {\bf A462} (2001) 152}\relax
\mciteBstWouldAddEndPuncttrue
\mciteSetBstMidEndSepPunct{\mcitedefaultmidpunct}
{\mcitedefaultendpunct}{\mcitedefaultseppunct}\relax
\EndOfBibitem
\bibitem{Golonka:2005pn}
P.~Golonka and Z.~Was, \ifthenelse{\boolean{articletitles}}{{\it {PHOTOS Monte
  Carlo: a precision tool for QED corrections in $Z$ and $W$ decays}},
  }{}\href{http://dx.doi.org/10.1140/epjc/s2005-02396-4}{Eur.\ Phys.\ J.\  {\bf
  C45} (2006) 97}, \href{http://arxiv.org/abs/hep-ph/0506026}{{\tt
  arXiv:hep-ph/0506026}}\relax
\mciteBstWouldAddEndPuncttrue
\mciteSetBstMidEndSepPunct{\mcitedefaultmidpunct}
{\mcitedefaultendpunct}{\mcitedefaultseppunct}\relax
\EndOfBibitem
\bibitem{Allison:2006ve}
GEANT4 collaboration, J.~Allison {\em et~al.},
  \ifthenelse{\boolean{articletitles}}{{\it {Geant4 developments and
  applications}}, }{}\href{http://dx.doi.org/10.1109/TNS.2006.869826}{IEEE
  Trans.\ Nucl.\ Sci.\  {\bf 53} (2006) 270}\relax
\mciteBstWouldAddEndPuncttrue
\mciteSetBstMidEndSepPunct{\mcitedefaultmidpunct}
{\mcitedefaultendpunct}{\mcitedefaultseppunct}\relax
\EndOfBibitem
\bibitem{Agostinelli:2002hh}
GEANT4 collaboration, S.~Agostinelli {\em et~al.},
  \ifthenelse{\boolean{articletitles}}{{\it {GEANT4: A simulation toolkit}},
  }{}\href{http://dx.doi.org/10.1016/S0168-9002(03)01368-8}{Nucl.\ Instrum.\
  Meth.\  {\bf A506} (2003) 250}\relax
\mciteBstWouldAddEndPuncttrue
\mciteSetBstMidEndSepPunct{\mcitedefaultmidpunct}
{\mcitedefaultendpunct}{\mcitedefaultseppunct}\relax
\EndOfBibitem
\bibitem{LHCb-PROC-2011-006}
M.~Clemencic {\em et~al.}, \ifthenelse{\boolean{articletitles}}{{\it {The \lhcb
  simulation application, \gauss: design, evolution and experience}},
  }{}\href{http://dx.doi.org/10.1088/1742-6596/331/3/032023}{{J.\ of Phys.\ :
  Conf.\ Ser.\ } {\bf 331} (2011) 032023}\relax
\mciteBstWouldAddEndPuncttrue
\mciteSetBstMidEndSepPunct{\mcitedefaultmidpunct}
{\mcitedefaultendpunct}{\mcitedefaultseppunct}\relax
\EndOfBibitem
\bibitem{Breiman}
L.~Breiman, J.~H. Friedman, R.~A. Olshen, and C.~J. Stone, {\em Classification
  and regression trees}, Wadsworth international group, Belmont, California,
  USA, 1984\relax
\mciteBstWouldAddEndPuncttrue
\mciteSetBstMidEndSepPunct{\mcitedefaultmidpunct}
{\mcitedefaultendpunct}{\mcitedefaultseppunct}\relax
\EndOfBibitem
\bibitem{delAmoSanchez:2010yp}
BaBar collaboration, P.~del Amo~Sanchez {\em et~al.},
  \ifthenelse{\boolean{articletitles}}{{\it {Dalitz plot analysis of $D_s^+ \to
  K^+ K^- \pi^+$}},
  }{}\href{http://dx.doi.org/10.1103/PhysRevD.83.052001}{Phys.\ Rev.\  {\bf
  D83} (2011) 052001}, \href{http://arxiv.org/abs/1011.4190}{{\tt
  arXiv:1011.4190}}\relax
\mciteBstWouldAddEndPuncttrue
\mciteSetBstMidEndSepPunct{\mcitedefaultmidpunct}
{\mcitedefaultendpunct}{\mcitedefaultseppunct}\relax
\EndOfBibitem
\bibitem{Bonvicini:2008jw}
CLEO collaboration, G.~Bonvicini {\em et~al.},
  \ifthenelse{\boolean{articletitles}}{{\it {Dalitz plot analysis of the $D^+
  \rightarrow K^- \pi^+ \pi^+$ decay}},
  }{}\href{http://dx.doi.org/10.1103/PhysRevD.78.052001}{Phys.\ Rev.\  {\bf
  D78} (2008) 052001}, \href{http://arxiv.org/abs/0802.4214}{{\tt
  arXiv:0802.4214}}\relax
\mciteBstWouldAddEndPuncttrue
\mciteSetBstMidEndSepPunct{\mcitedefaultmidpunct}
{\mcitedefaultendpunct}{\mcitedefaultseppunct}\relax
\EndOfBibitem
\bibitem{Skwarnicki:1986xj}
T.~Skwarnicki, {\em {A study of the radiative cascade transitions between the
  Upsilon-prime and Upsilon resonances}}, PhD thesis, Institute of Nuclear
  Physics, Krakow, 1986,
  {\href{http://inspirehep.net/record/230779/files/230779.pdf}{DESY-F31-86-02}%
}\relax
\mciteBstWouldAddEndPuncttrue
\mciteSetBstMidEndSepPunct{\mcitedefaultmidpunct}
{\mcitedefaultendpunct}{\mcitedefaultseppunct}\relax
\EndOfBibitem
\end{mcitethebibliography}

\end{document}